\documentclass[10pt, conference, compsocconf]{IEEEtran}

\usepackage[ruled, vlined, nofillcomment, linesnumbered]{algorithm2e}

\usepackage{cite}
\usepackage{url}
\usepackage[cmex10]{amsmath}
\usepackage{amssymb}
\usepackage{graphicx}
\usepackage[english]{babel}
\usepackage{booktabs}
\usepackage{fixltx2e}

\usepackage[font=footnotesize,caption=false]{subfig}

\usepackage{tikz}
\usetikzlibrary{arrows,shapes,positioning,fit,calc,matrix,trees}
\usepackage{pgfplots}

\usetikzlibrary{external}
\newcommand{\set}[1]{\left\{#1\right\}}
\newcommand{\pr}[1]{\left(#1\right)}
\newcommand{\fpr}[1]{\mathopen{}\left(#1\right)}

\newcommand{\fspr}[1]{\mathopen{}\left[#1\right]}

\newcommand{\abs}[1]{{\left|#1\right|}}

\newcommand{\enset}[2]{\left\{#1 ,\ldots , #2\right\}}
\newcommand{\enpr}[2]{\pr{#1 ,\ldots , #2}}
\newcommand{\enlst}[2]{{#1} ,\ldots , {#2}}

\newcommand{\real}{\mathbb{R}}

\newcommand{\funcdef}[3]{{#1}:{#2} \to {#3}}
\newcommand{\define}{\leftarrow}

\DeclareRobustCommand{\dispfunc}[2]{%
  \ensuremath{%
  \ifthenelse{\equal{#2}{}}%
    {\mathit{#1}}%
    {\mathit{#1}\fpr{#2}}}}

\newcommand{\supp}[1]{{\dispfunc{fr}{#1}}}
\newcommand{\cp}[1]{{\dispfunc{cp}{#1}}}
\newcommand{\rank}[1]{{\dispfunc{rnk}{#1}}}
\newcommand{\zind}[1]{{\dispfunc{z_{\textsc{ind}}}{#1}}}
\newcommand{\zpart}[1]{{\dispfunc{z_{\textsc{prt}}}{#1}}}
\newcommand{\indi}[1]{\mathit{I}\fspr{#1}}

\newcommand{\mean}[1]{\operatorname{E}\fspr{#1}}
\newcommand{\meanb}[1]{\operatorname{E}\mathopen{}\big[{#1}\big]}

\newcommand{\varb}[1]{\operatorname{Var}\mathopen{}\big[{#1}\big]}

\newtheorem{theorem}{Theorem}
\newtheorem{lemma}[theorem]{Lemma}
\newtheorem{proposition}[theorem]{Proposition}

\newtheorem{example}{Example}

\def\clap#1{\hbox to 0pt{\hss#1\hss}}

\def\mathrlap{\mathpalette\mathrlapinternal}
\def\mathrlapinternal#1#2{%
\rlap{$\mathsurround=0pt#1{#2}$}
}

\SetKwComment{tcpas}{\{}{\}}
\SetCommentSty{textnormal}
\SetArgSty{textnormal}
\SetKw{False}{false}
\SetKw{True}{true}
\SetKw{Null}{null}
\SetKwInOut{Output}{output}
\SetKwInOut{Input}{input}
\SetKw{AND}{and}
\SetKw{OR}{or}
\SetKw{Break}{break}

\pgfdeclarelayer{background}
\pgfdeclarelayer{foreground}
\pgfsetlayers{background,main,foreground}

\definecolor{yafaxiscolor}{rgb}{0.3, 0.3, 0.3}

\definecolor{yafcolor1}{rgb}{0.4, 0.165, 0.553}
\definecolor{yafcolor2}{rgb}{0.949, 0.482, 0.216}
\definecolor{yafcolor3}{rgb}{0.47, 0.549, 0.306}
\definecolor{yafcolor4}{rgb}{0.925, 0.165, 0.224}
\definecolor{yafcolor5}{rgb}{0.141, 0.345, 0.643}
\definecolor{yafcolor6}{rgb}{0.965, 0.933, 0.267}
\definecolor{yafcolor7}{rgb}{0.627, 0.118, 0.165}
\definecolor{yafcolor8}{rgb}{0.878, 0.475, 0.686}

\newlength{\yafaxispad}
\newlength{\yaftlpad}
\newlength{\yaflabelpad}
\newlength{\yafaxiswidth}
\newlength{\yafticklen}
\makeatletter
\def\pgfplots@drawtickgridlines@INSTALLCLIP@onorientedsurf#1{}
\makeatother

\newcommand{\yafdrawxaxis}[2]{
	\pgfplotstransformcoordinatex{#1}\let\xmincoord=\pgfmathresult 
	\pgfplotstransformcoordinatex{#2}\let\xmaxcoord=\pgfmathresult 
	\pgfsetlinewidth{\yafaxiswidth} 
	\pgfsetcolor{yafaxiscolor}
	\pgfpathmoveto{\pgfpointadd{\pgfpointadd{\pgfplotspointrelaxisxy{0}{0}}{\pgfqpointxy{\xmincoord}{0}}}{\pgfqpoint{-0.5\yafaxiswidth}{\yafaxispad}}}
	\pgfpathlineto{\pgfpointadd{\pgfpointadd{\pgfplotspointrelaxisxy{0}{0}}{\pgfqpointxy{\xmaxcoord}{0}}}{\pgfqpoint{0.5\yafaxiswidth}{\yafaxispad}}}
	\pgfusepath{stroke}

}
\newcommand{\yafdrawyaxis}[2]{
	\pgfplotstransformcoordinatey{#1}\let\ymincoord=\pgfmathresult 
	\pgfplotstransformcoordinatey{#2}\let\ymaxcoord=\pgfmathresult 
	\pgfsetlinewidth{\yafaxiswidth} 
	\pgfsetcolor{yafaxiscolor}
	\pgfpathmoveto{\pgfpointadd{\pgfpointadd{\pgfplotspointrelaxisxy{0}{0}}{\pgfqpointxy{0}{\ymincoord}}}{\pgfqpoint{\yafaxispad}{-0.5\yafaxiswidth}}}
	\pgfpathlineto{\pgfpointadd{\pgfpointadd{\pgfplotspointrelaxisxy{0}{0}}{\pgfqpointxy{0}{\ymaxcoord}}}{\pgfqpoint{\yafaxispad}{0.5\yafaxiswidth}}}
	\pgfusepath{stroke}
}

\newcommand{\yafdrawaxis}[4]{\yafdrawxaxis{#1}{#2}\yafdrawyaxis{#3}{#4}}

\pgfplotscreateplotcyclelist{yaf}{%
{yafcolor1,mark options={scale=0.75},mark=o}, 
{yafcolor2,mark options={scale=0.75},mark=square},
{yafcolor3,mark options={scale=0.75},mark=triangle},
{yafcolor4,mark options={scale=0.75},mark=o},
{yafcolor5,mark options={scale=0.75},mark=o},
{yafcolor6,mark options={scale=0.75},mark=o},
{yafcolor7,mark options={scale=0.75},mark=o},
{yafcolor8,mark options={scale=0.75},mark=o}} 

\pgfplotsset{axis y line=left, axis x line=bottom,
	tick align=outside,
	compat = 1.3,
	tickwidth=\yafticklen,
	clip = false,
	every axis title shift = 0pt,
    x axis line style= {-, line width = 0pt, opacity = 0},
    y axis line style= {-, line width = 0pt, opacity = 0},
    x tick style= {line width = \yafaxiswidth, color=yafaxiscolor, yshift = \yafaxispad},
    y tick style= {line width = \yafaxiswidth, color=yafaxiscolor, xshift = \yafaxispad},
    x tick label style = {font=\scriptsize, yshift = \yaftlpad},
    y tick label style = {font=\scriptsize, xshift = \yaftlpad},
    every axis y label/.style = {at = {(ticklabel cs:0.5)}, rotate=90, anchor=center, font=\scriptsize, yshift = -\yaflabelpad},
    every axis x label/.style = {at = {(ticklabel cs:0.5)}, anchor=center, font=\scriptsize, yshift = \yaflabelpad},
    x tick label style = {font=\scriptsize, yshift = 1pt},
    grid = major,
    major grid style  = {dash pattern = on 1pt off 3 pt},
	every axis plot post/.append style= {line width=\yafaxiswidth} ,
	legend cell align = left,
	legend style = {inner sep = 1pt, cells = {font=\scriptsize}},
	legend image code/.code={%
		\draw[mark repeat=2,mark phase=2,#1] 
		plot coordinates { (0cm,0cm) (0.15cm,0cm) (0.3cm,0cm) };%
	} 
}

\newlength{\adjlength}
\newcommand{\adjshort}[4]{
\pgfmathtruncatemacro{\adjside}{#1 - 1}
\pgfmathtruncatemacro{\adjend}{#1 + #2 + #3}
\fill[fill = #4, line width = 1pt] (0, #1\adjlength) -- (\adjside\adjlength, #1\adjlength)
\ifnum#2>0
\foreach \x in {1,..., #2} {-- ++(\adjlength, 0) -- ++(0, \adjlength) }
\fi
-- ++(0, #3\adjlength)  -- (0, \adjend\adjlength) -- cycle;
\fill[fill = #4, line width = 1pt] (#1\adjlength, 0) -- (#1\adjlength, \adjside\adjlength)
\ifnum#2>0
\foreach \x in {1,..., #2} {-- ++(0, \adjlength) -- ++(\adjlength, 0) }
\fi
-- ++(#3\adjlength, 0)  -- (\adjend\adjlength, 0) -- cycle;
}

\xspaceaddexceptions{$}

\makeatletter
\tikzset{circle split part fill/.style  args={#1,#2}{%
 alias=tmp@name, 
  postaction={%
    insert path={
     \pgfextra{%
     \pgfpointdiff{\pgfpointanchor{\pgf@node@name}{center}}%
                  {\pgfpointanchor{\pgf@node@name}{east}}%
     \pgfmathsetmacro\insiderad{\pgf@x}
      \fill[#1] (\pgf@node@name.base) ([xshift=-\pgflinewidth]\pgf@node@name.east) arc
                          (0:360:\insiderad-\pgflinewidth)--cycle;
      \fill[#2] (\pgf@node@name.base) ([xshift=\pgflinewidth]\pgf@node@name.west)  arc
                           (180:360:\insiderad-\pgflinewidth)--cycle;            
         }}}}}  
\makeatother   

\makeatletter
\def\closedcycley{%
    -| (perpendicular cs: 
        horizontal line through={(current plot begin)}, 
        vertical line through={(\pgfplots@ZERO@x,\pgfplots@ZERO@y)})
    -- cycle
}%
\makeatother

\begin{document}
\title{Itemsets for Real-valued Datasets}

\author{\IEEEauthorblockN{Nikolaj Tatti}
\IEEEauthorblockA{HIIT, Department of Information and Computer Science
Aalto University, Finland\\
nikolaj.tatti@aalto.fi}
}

\maketitle

\begin{abstract}
Pattern mining is one of the most well-studied subfields in exploratory data
analysis. While there is a significant amount of literature on how to discover
and rank itemsets efficiently from binary data, there is surprisingly little
research done in mining patterns from real-valued data. In this paper we
propose a family of quality scores for real-valued itemsets. We approach the
problem by considering casting the dataset into a binary data and computing the
support from this data. This naive approach requires us to select thresholds.
To remedy this, instead of selecting one set of thresholds, we treat thresholds
as random variables and compute the average support.  We show that we can
compute this support efficiently, and we also introduce two normalisations, namely
comparing the support against the independence assumption and, more generally,
against the partition assumption. Our experimental evaluation demonstrates that we can discover
statistically significant patterns efficiently.
\end{abstract}


%
\IEEEpeerreviewmaketitle

\section{Introduction}\label{sec:intro}
Pattern mining is one of the most well-studied subfields in exploratory data
analysis. While there is a significant amount of literature on how to discover
and rank itemsets efficiently from binary data, there is surprisingly little
research done in mining patterns from real-valued data. In this paper we
propose a family of quality scores for real-valued itemsets.

In order to motivate our approach,
assume that we are given a dataset $D$ containing real numbers and a miner for mining
itemsets from a binary data.  The most straightforward way to use the miner to
find patterns from $D$ is to transform $D$ into a binary data, and apply the miner.
More formally, assume that we have selected a threshold $t_i$ for every item
$i$ in the dataset. Then we define a binary data $B$ by setting $b_{ji} =
1$, if $d_{ji} \geq t_i$, and 0 otherwise, where $j$ ranges over all
transactions of $D$.

This approach has two immediate setbacks. Firstly, we have to select the
thresholds $t_i$. In addition, such a measure is coarse, any intricate
interaction between items is destroyed as data values are categorised into two
coarse categories, 0s and 1s.  Hence, instead of selecting just one set of
thresholds, we will \emph{vary} $t_i$, and instead of computing support only for one
dataset, we will compute an average support. More formally, we will attach a
distribution $p(R_i = t_i)$ to each threshold and compute the mean
$\mean{\supp{X; B}}$, where $\supp{X; B}$ is the frequency (support) of an
itemset $X$ in a binarized data $B$.

This approach has several benefits. First of all, the support is monotonically
decreasing, which allows us to discover all frequent itemsets efficiently.  On
the other hand, we will show that we can compute the support efficiently, even
though it involves taking an average over a complex function.

We still need to choose the threshold distribution $p(R_i = t_i)$. In this work
we focus on a specific distribution involved with copulas~\cite{nelsen:06:copula}: roughly
speaking, we will define $p(R_i \leq d_{ji}) = k/ (\abs{D} - 1)$, where $k$ is
the rank of the $j$th transaction after data is sorted w.r.t. the $i$th column.
We will see that this distribution induces a support in which the actual values
of individual items do not matter, instead the support is based on the ranks of the values.
Interestingly enough, several popular statistical tests, such as the Mann-Whitney U test or the Wilcoxon
signed-rank test, are also based on the ranks of values.



A standard technique in pattern mining is to compare the observed support against the
expected value under some null hypothesis, where the hypothesis is typically an
independence assumption. Here we consider two approaches, in the first approach
we do a $z$-normalisation by comparing the support against the independence
assumption.  In our second approach, we generalise the null hypothesis to a
partition model, where we assume that items from different parts of the partition
are independent. A particular difficulty with these approaches is that in order
to compute them we need to compute the expected mean and the variance.  While
this is trivial when dealing with simple transactional data, it becomes
intricate since the threshold distribution actually depends on the dataset.
Nevertheless, we can compute the exact mean and variance for the independence
assumption and exact mean and asymptotic variance for the partition assumption.
Interestingly enough, the independence test is non-parametric, that is the mean and the variance
depend only on the number of datapoints, whereas in the partition assumption 
we need to estimate parameters from the dataset.

The rest paper of the paper is organized as follows. We introduce preliminary
notation in Section~\ref{sec:prel}.  We define our general measure in
Section~\ref{sec:measure} and introduce copula support in
Section~\ref{sec:threshold}. We present an independence test in Section~\ref{sec:ind}
and test based on partitions in Section~\ref{sec:partition}.
We discuss related work in Section~\ref{sec:related} and present
our experiments in Section~\ref{sec:exp}. Finally, we conclude our paper with
remarks in Section~\ref{sec:conclusions}.

\section{Preliminaries and Notation}\label{sec:prel}
In this section we introduce the preliminary notation.

A \emph{dataset} $D$ is a multiset of $N$ transactions $d_1, \ldots, d_N$,
where $d_j \in \real^K$ is a vector of length $K$. We will often use $N = \abs{D}$ as
the number of datapoints and $K$ as the dimension of the dataset.
We treat each vector
$d_i$ as a sample from an unknown distribution, $p(a_1, \ldots, a_K)$.
We refer to the random variables $a_i$ as \emph{items}, or as \emph{features}.

Let $A = \enset{a_1}{a_K}$ be the set of all items.  An \emph{itemset} $X$ is a
set of items $X \subseteq A$. Assume that you are given an itemset $X$ and a
\emph{binary} vector $d \in \set{0, 1}^K$. We say that $d$ \emph{covers} $X$ if
$d_i = 1$, for every $a_i \in X$. We will use standard notation,
by writing $x_1\cdots x_M$ to mean $\enset{x_1}{x_M}$.

Assume now that we are given a collection of binary vectors $D = d_1,\ldots, d_N$.
We define the \emph{support} or the \emph{frequency} of an itemset $X$ as the proportion
of transactions in $D$ covering $X$,
\[
	\supp{X; D} = \frac{\abs{\set{1 \leq i \leq N; d_i \text{ covers } X}}}{N}\quad.
\]
An important property of the support is that it is monotonically decreasing,
that is, $\supp{X; D} \leq \supp{Y; D}$, if $Y \subseteq X$. This property
allows us to use efficient techniques~\cite{agrawal:96:fast} to discover \emph{all} itemsets
whose frequency is higher than some given threshold.

\section{Itemset support for real-valued data}\label{sec:measure}
In this section we define our measure for real-valued data. 
In order to do so,
let $D$ be a dataset over $K$ items, $\enlst{a_1}{a_K}$, and $N$ transactions.
Assume that we are given a threshold $t_i \in \real$ for each item $a_i$.  Let
us write $T = \enpr{t_1}{t_K}$. Given a vector $x \in \real^K$ of length $K$,
we define $y = x_T$ to be a \emph{binary} vector with $y_i = 1$ if $x_i \geq t_i$,
and $0$ otherwise.
We now define a \emph{binarized} data $D_T$ to be
\[
	D_T = \set{x_T \mid x \in D}\quad.
\]
Essentially, $D_T$ is a dataset where each value is binarized either to $0$
or to $1$, depending on the threshold. We can now compute a support for a given
itemset $X$ by computing $\supp{X; D_T}$.

The problem with this approach is that we need to select a threshold set $T$.
Additionally, once we have made this choice, the treatment of values in $D$ is
coarse: a value slightly higher than the threshold contributes to the support
as much as the values that are significantly higher.

To remedy this, we treat thresholds as random variables. That is, we have $K$
random variables, $\enlst{R_1}{R_K}$. We will assume that each threshold is
assigned independently, that is, $R_i$ are independent variables.  We will go
over some of the natural choices for distributions of $R_i$ later on.  If we
write $p(R_i = t_i)$ to be the density function of the $i$th threshold, we can now
define support as an average support, where the mean is taken over the possible
thresholds, that is,
\[
\begin{split}
	\supp{X; D, p} & =  \mean{\supp{X; D_T}} \\
	& = \int_{t_1} \cdots \int_{t_K} \supp{X; D_T} \prod_{i = 1}^K p(R_i = t_i)dt_i\quad.
\end{split}
\]

The important property of this support is that it is monotonically decreasing.
This allows us to mine all frequent itemsets using the standard pattern mining search.

\begin{proposition}
Assume two itemsets $X, Y$ such that $X \subseteq Y$.  Then $\supp{X; D, p} \geq \supp{Y; D, p}$.
\end{proposition}

\begin{IEEEproof}
For any given threshold set $T$, we have $\supp{X; D_T} \geq \supp{Y; D_T}$. It follows
immediately, that $\mean{\supp{X; D_T}} \geq \mean{\supp{Y; D_T}}$, which proves the proposition.
\end{IEEEproof}

Computing the support from the definition is awkward as it requires taking
$\abs{X}$ integrals.  Fortunately, we can rewrite the support in a much more accessible form. 

\begin{proposition}
Assume a dataset $D$ with $N$ transactions and a distribution $p$ over the thresholds.
Then the support of itemset $X$ is equal to
\[
	\supp{X; D, p} = \frac{1}{N} \sum_{x \in D} \prod_{i \in X} p(R_i \leq x_i)\quad.
\]

\end{proposition}
\begin{IEEEproof}
We can rewrite the support as
\[
\begin{split}
	\supp{X; D, p} & =  \mean{\supp{X; D_T}} \\
	               & =  \frac{1}{N}\sum_{x \in D} p(x_T \text{ covers } X)\quad. \\
\end{split}
\]
Transaction $y = x_T$ covers $X$ if only if $y_i \geq R_i$ for each $i \in X$.
Since $R_i$ are independent, it follows that
\[
	p(x_T \text{ covers } X) =  \prod_{i \in X} p(R_i \leq x_i)\quad.
\]
This completes the proof.
\end{IEEEproof}

\section{Copula Support}\label{sec:threshold}
Our measure depends on the threshold distribution. In this section we focus on a specific
distribution related to copulas.



Assume that we are given a dataset $D = d_1, \ldots, d_N$.  Let us assume for
simplicity that for each item, say $a_j$, the data points $d_{ij}$ are unique.  Fix an item $a_j$ and
for notational simplicity let us assume that the datapoints are ordered according to the $j$th
item, $d_{ij} < d_{(i + 1)j}$ for $i = 1, \ldots, N - 1$.
Let us define the probability of a threshold
$R_i$ by requiring that the threshold will hit the interval $[d_{ij}, d_{(i
+ 1)j}]$ with a probability of $1 / (N - 1)$, where $i = 1, \ldots, N - 1$.
In other words, the cumulative distribution is equal to
\[
	p(R_j < d_{ij}) = \frac{i - 1}{N - 1}\quad.
\]
This gives us straightforward way of computing the support. Given a dataset $D$
of $N$ points, we compute $r_{ij} = (c - 1) / (N - 1)$, where $c$
is the rank of the $i$th transaction according to the $j$th column. We can now define
a \emph{copula}\footnote{Copula stands for a cumulative joint distribution of
random variables that have gone through such a transformation~\cite{nelsen:06:copula}.} support by
\[
	\cp{X; D} = \frac{1}{N}\sum_{i = 1}^N \rank{i; X, D},
\]
where $\rank{i; X, D} = \prod_{j \in X} r_{ij}$.

\begin{example}
Consider that we are given a dataset with 4 items and 3 transactions
\[
	  \set{(1.2, 4.5,  3.8, 8.9), (4.4,    4.7,    1.9,    8.8), (8.2,    8.5,    3.0,    6.5)}\ .
\]
The corresponding ranks $\set{r_{ij}}$ are then
\[
	  \set{(0, 0,  1, 1), (0.5,    0.5,    0,   0.5), (1,   1,    0.5,    0)}\ .
\]
For example, the copula support for $\set{a_2a_3}$ is then
\[
	\cp{a_2a_3} = \frac{1}{3}(0\times 1 + 0.5 \times 0 + 1 \times 0.5) = \frac{1}{6}\quad.
\]
\end{example}

As we see in the experiments, using $\cp{X, D}$ as a filtering condition is not enough.
Consequently, we also define $\cp{X; D, \alpha}$ by setting
\[
	p(R_i < d_i) = \max\pr{\min\pr{\frac{i - 1 - M}{N - 1 - 2M}, 1}, 0},
\]
where $M = \lfloor\alpha N\rfloor$, that is, the top $\alpha N$ items will be
always above threshold and the bottom $\alpha N$ will be always below threshold.

Copula support has some peculiar features. First of all, the support does not
depend on the actual values of $D$, only on their ranks. This makes this support
excellent for cases where computing the difference between the values of $D$
does not make sense.  In addition to that $\cp{a_i; D} = 1/2$
for any item, hence the support is not useful for selecting itemsets of size
$1$. Even though, we assume that $D$ has independent samples, the ranks $r_{ij}$
are no longer independent. However, if we assume independence between the items,
we can compute the mean and the variance as we will see in the next section.


\section{Copula support as a statistical test}\label{sec:ind}
A standard technique in pattern mining is to compare the observed support
against the independence model. In this section we demonstrate how to do
this comparison for copula support. More specifically, we are interested in
the quantity
\[
	\zind{X; D} = \sqrt{N} \frac{\cp{X; D} - \mu}{\sigma},
\]
where $\mu$ and $\sigma$ are the mean and the variance of the copula support
under the null hypothesis.

We will now show how to compute the mean and the variance of the copula support. In fact,
if we set $M = \abs{X}$, then we will show that $\mu = 1/2^M$ and
\[
	\sigma^2 = \frac{(2N - 1)^M}{6^M(N - 1)^M} + \frac{(N - 2)^M(3N - 1)^M}{12^M(N - 1)^{2M - 1}} - \frac{N}{4^M}\quad.
\]
We will also
show that $\zind{X; D}$ approaches the Gaussian distribution $N(0, 1)$ as the number of data
points goes to infinity. 

To simplify the analysis we will make an assumption that the probability of a
tie between two values of an item is $0$. This assumption is reasonable if the dataset is
generated for example from sensor readings.



We will dedicate the remaining section to proving these results.  Note that
we cannot use Central Limit Theorem to prove the normality because the ranks of
individual rows are not independent. Case in point, $\cp{x}$ for a single item
will always be $1/2$, hence the variance will be $0$ for this case. 

In order to prove the result, we will first need to establish some notation.
Assume that we have $N$ samples, independent and identically distributed random variables,
$\mathcal{Y} = Y_1,\ldots,Y_N$, each sample is a vector of size $K$. Define
\[
	S_{ij} = \rank{i; j, \mathcal{Y}} = \frac{1}{N - 1}\sum_{k = 1}^N \indi{Y_{ij} > Y_{kj}},
\]
where $\indi{B}$ returns 1 if the statement $B$ is true, and $0$ otherwise.
Note that the term $\indi{Y_{ij} = Y_{ij}} = 0$, however, we keep it in the sum for notational
convenience. Similarly, we can now define 
\[
	U = \cp{X; \mathcal{Y}} = \frac{1}{N} \sum_{i = 1}^N \prod_{j \in X} S_{ij}\quad.
\]
If we are given a dataset $D$, then $\cp{X; D}$ is an estimate of the random variable $U$.
Our goal is to compute $\mu = \mean{U}$ and $\sigma^2 = \varb{\sqrt{N}U}$.

Note that since we assume that $Y_{ij}$ and $Y_{kl}$ are independent for $j
\neq l$, it follows also that $S_{ij}$ and $S_{kl}$ are also independent for $j
\neq l$. However, unlike $Y_{ij}$ and $Y_{kj}$, $S_{ij}$ and $S_{kj}$ are
\emph{not} independent.

In order to continue we need the following lemma.
\begin{lemma}
\label{lem:probs}
Fix $j$ and let $i$, $k$, and $l$ be distinct integers.
Then
\[
\begin{split}
	p(Y_{ij} > Y_{kj}) &= 1/2,\\
	p(Y_{ij} > Y_{lj}, Y_{kj} > Y_{lj}) &= 1/3,\\
	p(Y_{ij} > Y_{kj}, Y_{kj} > Y_{lj}) &= 1/6\quad.
\end{split}
\]

\end{lemma}

\begin{IEEEproof}
Since the probability of a having a tie between variables is $0$, using the
symmetry argument, the probability $Y_{ij}$ will be larger than $Y_{kj}$ is
$1/2$.

Similarly, if we sort the three variables based on their value, there are 6
possible permutations, each permutation has a probability of $1/6$.  There are
two permutations that satisfy the second event, namely $Y_{ij} > Y_{kj} >
Y_{lj}$ and $Y_{kj} > Y_{ij} > Y_{lj}$. This shows that the probability of the
second event is equal to $1/3$. Finally, there is only one permutation
that satisfies the third event, namely, $Y_{ij} > Y_{kj} >
Y_{lj}$, which proves the lemma.
\end{IEEEproof}

We will first compute the mean of $U$.

\begin{proposition}
The average of $U$ is $\mean{U} = 1/2^M$.
\end{proposition}
\begin{IEEEproof}
According to Lemma~\ref{lem:probs}, $\mean{S_{ij}} = 1/2$.
Since $S_{ij}$ and $S_{kl}$ are independent for $j \neq l$, we can write 
\[
	\mean{U} =  \frac{1}{N} \sum_{i = 1}^N \prod_{j \in X} \mean{S_{ij}} = \frac{1}{N} \sum_{i = 1}^N \prod_{j \in X} \frac{1}{2} = \frac{1}{2^M}\quad.
\]
\end{IEEEproof}
This proves the result.

Our next step is to compute the variance of $U$.  Since the variables $S_{ij}$
are not independent, we will have to compute them in two stages. Our first step
is to compute the second moment of $S_{ij}$.

\begin{lemma}
\label{lem:second}
The second moment of $S_{ij}$ is equal to
\[
	\mean{S_{ij}^2} = \frac{2N - 1}{6(N - 1)}\quad.
\]
\end{lemma}

\begin{IEEEproof}
Decompose the second moment into two sums,
\[
\begin{split}
	\mean{S_{ij}^2}  & = \frac{1}{(N - 1)^2} \meanb{\big(\sum_{k \neq i} \indi{Y_{ij} > Y_{kj}}\big)^2} \\
	& =\frac{1}{(N - 1)^2}\sum_{k \neq i} p(Y_{ij} > Y_{kj}) \\
	&\qquad+ \frac{1}{(N - 1)^2}\sum_{k \neq i} \sum_{l \neq k, i} p(Y_{ij} > Y_{kj}, Y_{ij} > Y_{lj})\ .
\end{split}
\]
According to Lemma~\ref{lem:probs}, the terms in the first sum are equal to $1/2$ while
the terms in the second sum are equal to $1/3$. This gives us
\[
\begin{split}
	\mean{S_{ij}^2} & = \frac{1}{(N - 1)^2}\big((N - 1)/2 + (N - 1)(N - 2)/3\big) \\
	                & = \frac{1}{6(N - 1)}(3 + 2(N - 2)) = \frac{2N - 1}{6(N - 1)}\quad.  \\
\end{split}
\]
This completes the proof.
\end{IEEEproof}

Our next step is to compute the cross-moment of $S_{ij}$.

\begin{lemma}
\label{lem:cross}
The cross-moment is equal to
\[
	\mean{S_{ij}S_{kj}} = \frac{(N - 2)(3N - 1)}{12(N - 1)^2}\quad.
\]
\end{lemma}

\begin{IEEEproof}
Decompose the moment into four sums
\[
\begin{split}
	 &\mean{S_{ij}S_{kj}} \\
	 &\  = \frac{1}{(N - 1)^2}\meanb{\big(\sum_{m \neq i} \indi{Y_{ij} > Y_{mj}}\big)\big(\sum_{n \neq k} \indi{Y_{kj} > Y_{nj}}\big)} \\
	 &\  = \frac{A + B + C + D}{(N - 1)^2},
\end{split}
\]
where
\[
\begin{split}
	A & = \sum_{m \neq i, k} p(Y_{ij} > Y_{mj}, Y_{kj} > Y_{mj}), \\
	B & = \sum_{m \neq i, k} \sum_{n \neq m, i, k} p(Y_{ij} > Y_{mj}, Y_{kj} > Y_{nj}), \\
	C & = \sum_{m \neq i, k} p(Y_{ij} > Y_{kj}, Y_{kj} > Y_{nj}), \quad\text{and} \\
	D & = \sum_{m \neq i, k} p(Y_{ij} > Y_{mj}, Y_{kj} > Y_{ij})\quad. 
\end{split}
\]
The random variables in the term of the sum of $B$ are all independent, hence
the probability is equal to $1/4$.  According to Lemma~\ref{lem:probs} the term
in the sum of $A$ is equal to $1/3$ and the term in the sum for $C$ and $D$ is
equal to $1/6$. This gives us
\[
	A = \frac{N - 2}{3},\  B = \frac{(N - 2)(N - 3)}{4},\  C = D = \frac{N - 2}{6}\quad.
\]
Grouping the terms gives us
\[
\begin{split}
	\mean{S_{ij}S_{kj}} & =  \frac{4(N - 2) + 3(N - 2)(N - 3) + 4(N - 2)}{12(N - 1)^2} \\
	                    & =  \frac{(N - 2)(3N - 1)}{12(N - 1)^2}\quad. \\
\end{split}
\]
This completes the proof.
\end{IEEEproof}

We can now use both lemmas in order to compute the variance.
\begin{proposition}
\label{prop:indvariance}
The variance $\varb{\sqrt{N}U}$ is equal to
\[
	\sigma^2 = \frac{(2N - 1)^M}{6^M(N - 1)^M} + \frac{(N - 2)^M(3N - 1)^M}{12^M(N - 1)^{2M - 1}} - \frac{N}{4^M}\quad.
\]
\end{proposition}

\begin{IEEEproof}
We begin by splitting $\meanb{(\sqrt{N}U)^2}$ into two sums and applying Lemma~\ref{lem:second} and Lemma~\ref{lem:cross},
\[
\begin{split}
	\meanb{(\sqrt{N}U)^2} & = \frac{1}{N} \big( \sum_{i = 1}^N \prod_{j \in X} \mean{S_{ij}^2} + \sum_{i, k \atop i \neq k} \prod_{j \in X} \mean{S_{ij}S_{kj}} \big) \\
               &  = \frac{(2N - 1)^M}{6^M(N - 1)^M} + \frac{(N - 2)^M(3N - 1)^M}{12^M(N - 1)^{2M - 1}}\ .
\end{split}
\]
We can now use this to express the variance as
\[
\begin{split}
	\sigma^2 &= \meanb{(\sqrt{N}U)^2} - N\mean{U}^2\\
	& = \frac{(2N - 1)^M}{6^M(N - 1)^M} + \frac{(N - 2)^M(3N - 1)^M}{12^M(N - 1)^{2M - 1}} - \frac{N}{4^M}\quad.
\end{split}
\]


This proves the result.
\end{IEEEproof}

Finally, we show that $\zind{X; \mathcal{Y}}$ approaches a Gaussian distribution.
Note that this result does not depend on the assumption that items are independent.
Hence, we will be able to use the same result in the next section.

\begin{proposition}
\label{prop:normality}
The quantity $\sqrt{N}(U - \mean{U})$ approaches a Gaussian distribution
as $N$ approaches infinity.
\end{proposition}
We postpone the proof of this proposition to Appendix.

\section{Productive Itemsets and Copula Support}\label{sec:partition}

In the previous section we tested the support against the independence assumption.
A natural extension of this is to assume a
partition of the given itemset such that items are independent only when
they belong to different blocks of the partition. In fact, an approach suggested
in~\cite{DBLP:journals/tkdd/Webb10} mines itemsets from binary data whose
support is substantially larger than the expectation given by the partition.
In order to mimic this for real-valued data, we define
\[
	\zpart{X, P; D} = \frac{\cp{X; D} - \mu}{\sigma},
\]
where $P$ is a partition of $X$ and where $\mu$ and $\sigma$ is the mean and
the variance under the assumption that items belonging to different blocks in $P$ are
independent.  Our final goal is to find a partition that produces the lowest
score, that is, a partition that explains the support the best, $\zpart{X; D} =
\min_P  \zpart{X, P; D}$, where $P$ goes over all partitions of at least size
$2$. Note that we are only interested in one-side test. However, we can easily
adjust the formula for a symmetrical two-side test.  In addition,
in~\cite{DBLP:journals/tkdd/Webb10} the authors were looking only at partitions
of size $2$, whereas we go over all non-trivial partitions. 

In this section we show how we can compute the needed mean and the variance in
order to normalise the support. Unlike with the independence model, the test is no
longer non-parametric and we will have to estimate several parameters for each
subitemset in the partition. Moreover, we will only provide the variance only
when $N$ approaches infinity as the interactions between variables are complex
and hard to compute exactly for finite $N$.

We proceed as follows: We will first show what statistics
we need from each subitemset and how to compute them. Then we will show how to use
these statistics in order to compute the mean and the variance. 

\subsection{Statistics needed to compute the rank}
Assume that we are given an itemset $X = x_1 \cdots x_M$. This itemset will
eventually be a block in the partition.  Let $\mathcal{Y} = Y_1, \ldots, Y_N$ be $N$ data
samples.  Let us shorten $O_{ijx} = \frac{1}{N - 1}\indi{Y_{ix} > Y_{jx}}$.
Let us define
\[
	T_i = \rank{i; X, \mathcal{Y}} = \prod_{x \in X}  \sum_{j = 1}^N O_{ijx},
\]
which is essentially a product of normalised ranks of the $i$th datapoint.
Similar to Section~\ref{sec:ind}, let $U = \frac{1}{N}\sum_{i  = 1}^N T_i$, a random
variable corresponding to the copula support $\cp{X}$.

Ultimately, we will need three statistics from $X$, namely $\mu = \mean{U}$,
$\alpha = \meanb{T_1^2}$, and $\beta = \varb{\sqrt{N}U}$.
We will discuss how to estimate these statistics in the next subsection.
If $T_i$ were distributed independently, then $\beta = \alpha - \mu^2$.
However, $T_i$ are dependent. Fortunately, we know enough about the dependency
so that we can compute $\beta$.

In order to compute $\beta$ we need to introduce several random variables.
Let 
\[
	T_{ix} = \rank{i, X \setminus \set{x}, \mathcal{Y}} = \prod_{y \in X, y \neq x} \sum_{j = 1}^N O_{ijy}
\]
be the rank of the $i$th transaction for an itemset $X \setminus \set{x}$.  In
addition, let us define $C_{kx} = \sum_{i = 1}^N T_{ix}O_{ikx}$.  We can
express the variance $\beta$ with $\alpha$, $\mu$ and $C_{kx}$.  The benefit of
this is that we can estimate these parameters, and by doing so estimate
$\beta$, as we will demonstrate in the next subsection.

\begin{proposition}
\label{prop:partvar}
The variance $\beta$ approaches 
\[
	\alpha - (M + 1)^2\mu^2 + \frac{2}{N}\sum_{k = 1}^N\meanb{\big(\sum_{x \in X} C_{kx}\big)^2}
\]
as $N$ approaches infinity.
\end{proposition}

We postpone the proof of this proposition to Appendix.

\subsection{Estimating statistics} 

Unlike with $\zind{X}$, the mean and the variance of $\zind{X; P}$ depend on
the underlying distribution, and we are forced to estimate the statistics,
namely $\alpha$, $\beta$, $\mu$ described in the previous section. These
estimates are given in Algorithm~\ref{alg:estimate}. Estimating $\mu$ and
$\alpha$ is trivial. However, estimating $\beta$ is more intricate due to the
last term given in Proposition~\ref{prop:partvar}.

Assume that we are given a dataset $D$ and itemset $X$.  Fix $x \in X$ and
assume that $D$ is sorted based on $x$th column, largest first.
Let $Z = X \setminus \set{x}$. Note that $\rank{k; Z, D}$ is an estimate for $T_{kx}$.
Hence, we can estimate $C_{kx}$ as
\[
\begin{split}
	c_{kx} &= \frac{1}{N - 1}\sum_{i = 1}^{k - 1} \rank{i; Z, D} \\
	& = c_{(k - 1)x} + \frac{1}{N - 1} \rank{k - 1; Z, D}\quad.
\end{split}
\]
We can use the right-hand side to compute $c_{kl}$ for every $k$ efficiently,
and then use $c_{kl}$ to estimate $\beta$.  We can assume that we have
precomputed the order w.r.t. each item $x_l$ before the actual mining. Hence, the
cost of estimating the parameters is $O(N\abs{X})$. 


\begin{algorithm}[ht!]
\caption{\textsc{Estimate}, estimates the statistics needed for $\zpart{}$.}
\label{alg:estimate}
\Input{dataset $D$, itemset $X$}
\Output{estimates $\mu$, $\alpha$, and $\beta$}
$\mu \define \cp{X; D} $\;
$\alpha \define \frac{1}{N}\sum_{i = 1}^N \rank{i; X, D}^2$\; 
$c_{ix} \define 0$,\quad $i = 1,\ldots,N$,\quad $x \in X$\;

\ForEach {$x \in X$} {
	sort $D$ according to $x$, largest first\;
	\ForEach {$k \in [2, N]$} {
		$c_{kx} \define c_{(k - 1)x} + \frac{1}{N - 1}\rank{k - 1, X \setminus \set{x}, D}$\; 
	}
}
$\beta \define \alpha - (\abs{X} + 1)^2\mu^2 + \frac{2}{N} \sum_{k = 1}^N \big(\sum_{x \in X}c_{kx}\big)^2$\;
\Return{$\mu, \alpha, \beta$}\;
\end{algorithm}

We should stress that we use the same dataset to compute the estimates and to
compute $\zpart{X; D}$. This means that $\zpart{X, P; D}$ will be somewhat
skewed and we cannot interpret $\zpart{X, P; D}$ as a $p$-value. However, our
main goal is not to interpret the obtained values as a statistical test, rather
our goal is to rank patterns. 

\subsection{Computing z-score} 

Now that we have computed statistics for each itemset occurring in a partition,
we can combine them in order to compute the mean and the variance needed for
$\zpart{X}$.

\begin{proposition}
\label{prop:partstats}
Assume that we are given an itemset $X$ and a partition $P_1, \ldots, P_L$ of
$X$.  Let $\mathcal{Y} = Y_1, \ldots, Y_N$ be $N$ random data points.  Let $U =
\cp{X; \mathcal{Y}}$, and let $U_i = \cp{P_i; \mathcal{Y}}$.  Let $\mu_i =
\mean{U_i}$, $\alpha_i = \meanb{\rank{1; P_i, \mathcal{Y}}^2}$, $\beta_i = \lim_{N \to \infty}
\varb{\sqrt{N}U_i}$. 

Under the assumption that $P_i$ are independent, we have $\mean{U} = \prod_{i = 1}^L \mu_i$ and
\[
	\varb{\sqrt{N}U} \to \prod_{i = 1}^L \alpha_i  + (L - 1)\mu^2 +  \mu^2\sum_{i = 1}^L \frac{\beta_i - \alpha_i}{\mu_i^2}
\]
as $N$ approaches infinity.
\end{proposition}
We postpone the proof of this proposition to Appendix.

\section{Related Work}\label{sec:related}

While pattern mining has been well researched for binary data, the problem of
discovering patterns from real-valued data is open. The most straightforward
approach to mine patterns is to discretize data using threshold, see for example~\cite{DBLP:conf/sigmod/SrikantA96}.
Among methods that do not use thresholds,
Calders et
al.~\cite{DBLP:conf/kdd/CaldersGJ06} proposed 3 quality measures for itemsets
from numerical attributes. The first two measures were based on the extrema
values of the items in an itemset.  The most related measure to our work is the
third measure, $\mathit{supp}_\tau$, which is a generalisation of Kendall's
$\tau$, essentially the number of pairs in which all items are concordant.
Interestingly enough, similar to the copula support, $\mathit{supp}_\tau$ also
depends only the order of values not on the actual values. In this work we were
able to define two normalisations $\zind{X}$ and $\zpart{X}$ for our approach,
while the authors did not introduce any statistical normalisation for  $\mathit{supp}_\tau$.
We conjecture that a similar normalisation can be done also for $\mathit{supp}_\tau$.

Jaroszewicz and Korzen~\cite{DBLP:conf/sdm/JaroszewiczK07} suggested
discovering polynomial itemsets, essentially cross-moments from real-valued
data. We can show that for a certain threshold distribution, our support is equal to
the support of polynomial itemsets. Steinbach et
al.~\cite{DBLP:conf/kdd/SteinbachTXK04} considered several support functions
for itemsets, such as, taking the smallest value in a transaction among the items in the itemset. 

Ranking and filtering patterns based on a statistical test has been well
studied.  Brin et al. compared likelihood-ratio against independence
assumption~\cite{brin:97:beyond}. Webb proposed, among many other criteria, to
compare the observed support to an expected support of a partition of size
$2$ that fits best~\cite{DBLP:journals/tkdd/Webb10}. More complex null
hypotheses such as Bayesian networks~\cite{jaroszewicz:04:interestingness} or
Maximum Entropy models~\cite{tatti:08:maximum} have been also suggested.

Our approach has similarities with mining itemsets from uncertain
data~\cite{DBLP:conf/pakdd/ChuiKH07}, where instead of binary data, we have
real-valued values between $[0, 1]$ expressing the likelihood of the entry
being equal to 1. In fact, if we interpret $r_{ij}$ values computed in
Section~\ref{sec:threshold} as probabilistic dataset, then $\cp{X}$ will be the
same as the expected support computed from probabilistic dataset. However, in
probabilistic setting the entries are assumed to be independent, whereas in our
case they have an intricate dependency. Consequently, the variance given by
Propositions~\ref{prop:indvariance}~and~\ref{prop:partvar} do not hold for probabilistic datasets.  In
addition, we cannot compute frequentness measure suggested
by Bernecker et al.~\cite{DBLP:conf/kdd/BerneckerKRVZ09} in our case,
however we can estimate it by a normal distribution as suggested
by Calders et al.~\cite{DBLP:conf/icdm/CaldersGG10}.

Defining and computing a quality score for two real-valued variables,
essentially an itemset of length 2, is a surprisingly open problem. The
approach based on Information Theory was suggested in~\cite{reshef:01:novel}.
An interesting starting point is also a measure of concordance, see Definition
5.1.7~in~\cite{nelsen:06:copula}.  These approaches are suitable only for
itemsets of size 2 whereas we are interested in measuring the quality of
itemset of any size.  Finally, Sze\'{e}kely and
Rizzo~\cite{szekely:09:distance} suggested a measure based on how pair-wise
distances correlate. This measure is symmetric while our measure was specifically
designed to focus on large values.

\section{Experiments}\label{sec:exp}
In this secion we present our experiments.

\emph{Datasets:} We used 2 synthetic and 3 real-world data sets as our
benchmark data. The first dataset \emph{Ind} consists of $10\,000$ data points,
each of $100$ items, generated independently uniformly from the interval $[0,
1]$. The second dataset \emph{Plant} has the same dimensions as the first
dataset. In this dataset we planted 5 subspace clusters each having $4$ items:
We generated independently $5\times 10\,000$ boolean variables $B_{ti}$
indicating whether a transaction $t$ belongs to the $i$th cluster, a
transaction can belong to multiple clusters. We set $p(B_{ti} = 1) = 0.4$.  If
$B_{ti} = 1$, then we set the corresponding items to $0.5$. All other values
were set to $0$.  Finally, we added noise sampled uniformly from $[0, 1]$ .
As real-world benchmark datasets we used the following 3 gene expression data
sets: \emph{Alon}~\cite{alon:99:pnas}, Arabidopsis thaliana or \emph{Thalia},
and Saccharomyces cerevisiae or \emph{Yeast}.\!\footnote{\emph{Thalia} and
\emph{Yeast} are available at \url{http://www.tik.ee.ethz.ch/~sop/bimax/}}
The sizes of the datasets are given in Table~\ref{tab:basic}.

\begin{table}[t!]
\caption{Basic statistics of datasets and experiments}
\label{tab:basic}
\begin{tabular*}{\columnwidth}{@{\extracolsep{\fill}}lrrrr}
\toprule
Name & Size & Threshold & Time & $\abs{\text{patterns}}$ \\
\midrule
\emph{Ind} & $10\,000\times 100$ & $0.1$ & $7$m$37$s & $166\,750$ \\
\emph{Plant} & $10\,000\times 100$ & $0.1$ & $7$m$14$s & $171\,303$ \\[1mm]
\emph{Alon} & $2000\times 62$ & $0.26$ & $8$m$17$s & $393\,683$\\
\emph{Thalia} & $734\times 69$ & $0.12$ & $2$m$10$s & $148\,334$\\
\emph{Yeast} & $2993\times 173$ & $0.2$ & $19$m$50$s & $529\,872$\\
\bottomrule
\end{tabular*}
\end{table}

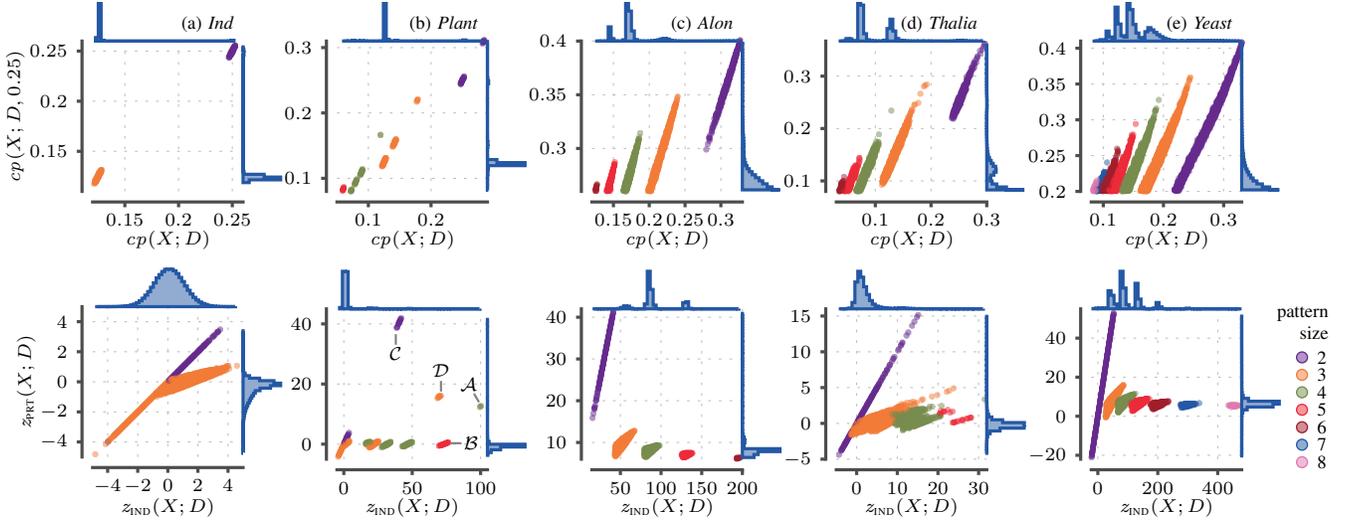
\begin{figure*}[ht!]
\newlength{\imgsize}
\setlength{\imgsize}{2cm}
\setlength{\tabcolsep}{-3pt}
\begin{tabular}{rrrrrr}
\begin{tikzpicture}
\pgfmathsetmacro{\imgwidth}{\imgsize}
\pgfmathsetmacro{\imgheight}{\imgsize}
\pgfmathsetmacro{\dxmin}{0.12}
\pgfmathsetmacro{\dxmax}{0.26}
\pgfmathsetmacro{\dymin}{0.11}
\pgfmathsetmacro{\dymax}{0.26}

\begin{axis}[xlabel={$\cp{X; D}$}, ylabel= {$\cp{X; D, 0.25}$}, title = {\scriptsize(a) \emph{Ind}}, title style = {anchor = south east, at = {(1, 1)}},
	name = scatter,
    width = \imgwidth,
    height = \imgheight,
	scale only axis,
    cycle list name=yaf,
	scatter/classes = {2={yafcolor1},3={yafcolor2},4={yafcolor3},5={yafcolor4},6={yafcolor7}},
	xmin = \dxmin, xmax = \dxmax, ymin = \dymin, ymax = \dymax,
    ]
\addplot[scatter, scatter src = explicit symbolic, only marks, mark size = 1, opacity = 0.5, mark = *, mark options = {line width = 0pt}] table[x index = 0, y index = 1, meta index = 2, header = false]
{ind_0.1_0.25_supmean.sct};
\pgfplotsextra{\yafdrawaxis{\dxmin}{\dxmax}{\dymin}{\dymax}}
\end{axis}
\begin{axis}[at = {(scatter.north west)},
    width = \imgwidth,
    height = 0.5cm,
	scale only axis,
	xmin = \dxmin, xmax = \dxmax,
	ytick = \empty, xtick = \empty, scaled ticks = false,
    ]
\addplot+[const plot, no markers, yafcolor5, fill = yafcolor5!50] table[x index = 1, y index = 0, header = false] {ind_0.1_0.25_supmean.hst} \closedcycle;
\end{axis}
\begin{axis}[at = {(scatter.south east)},
    height = \imgheight,
    width = 0.5cm,
	scale only axis,
	ymin = \dymin, ymax = \dymax,
	ytick = \empty, xtick = \empty, scaled ticks = false,
    ]
\addplot+[const plot, no markers, yafcolor5, fill = yafcolor5!50] table[x index = 2, y index = 3, header = false] {ind_0.1_0.25_supmean.hst} \closedcycley;
\end{axis}

\end{tikzpicture} &
\begin{tikzpicture}
\pgfmathsetmacro{\imgwidth}{\imgsize}
\pgfmathsetmacro{\imgheight}{\imgsize}
\pgfmathsetmacro{\dxmin}{0.05}
\pgfmathsetmacro{\dxmax}{0.29}
\pgfmathsetmacro{\dymin}{0.08}
\pgfmathsetmacro{\dymax}{0.31}

\begin{axis}[xlabel={$\cp{X; D}$}, title = {\scriptsize(b) \emph{Plant}}, title style = {anchor = south east, at = {(1, 1)}}, 
	name = scatter,
    width = \imgwidth,
    height = \imgheight,
	scale only axis,
    cycle list name=yaf,
	scatter/classes = {2={yafcolor1},3={yafcolor2},4={yafcolor3},5={yafcolor4},6={yafcolor7}},
	xmin = \dxmin, xmax = \dxmax, ymin = \dymin, ymax = \dymax,
    ]
\addplot[scatter, scatter src = explicit symbolic, only marks, mark size = 1, opacity = 0.5, mark = *, mark options = {line width = 0pt}] table[x index = 0, y index = 1, meta index = 2, header = false]
{plant_0.08_0.25_supmean.sct};
\pgfplotsextra{\yafdrawaxis{\dxmin}{\dxmax}{\dymin}{\dymax}}
\end{axis}
\begin{axis}[at = {(scatter.north west)},
    width = \imgwidth,
    height = 0.5cm,
	scale only axis,
	xmin = \dxmin, xmax = \dxmax,
	ytick = \empty, xtick = \empty, scaled ticks = false,
    ]
\addplot+[const plot, no markers, yafcolor5, fill = yafcolor5!50] table[x index = 1, y index = 0, header = false] {plant_0.08_0.25_supmean.hst} \closedcycle;
\end{axis}
\begin{axis}[at = {(scatter.south east)},
    height = \imgheight,
    width = 0.5cm,
	scale only axis,
	ymin = \dymin, ymax = \dymax,
	ytick = \empty, xtick = \empty, scaled ticks = false,
    ]
\addplot+[const plot, no markers, yafcolor5, fill = yafcolor5!50] table[x index = 2, y index = 3, header = false] {plant_0.08_0.25_supmean.hst} \closedcycley;
\end{axis}

\end{tikzpicture} &
\begin{tikzpicture}
\pgfmathsetmacro{\imgwidth}{\imgsize}
\pgfmathsetmacro{\imgheight}{\imgsize}
\pgfmathsetmacro{\dxmin}{0.12}
\pgfmathsetmacro{\dxmax}{0.33}
\pgfmathsetmacro{\dymin}{0.26}
\pgfmathsetmacro{\dymax}{0.4}

\begin{axis}[xlabel={$\cp{X; D}$}, title = {\scriptsize(c) \emph{Alon}}, title style = {anchor = south east, at = {(1, 1)}}, 
	name = scatter,
    width = \imgwidth,
    height = \imgheight,
	scale only axis,
    cycle list name=yaf,
	scatter/classes = {2={yafcolor1},3={yafcolor2},4={yafcolor3},5={yafcolor4},6={yafcolor7}},
	xmin = \dxmin, xmax = \dxmax, ymin = \dymin, ymax = \dymax,
    ]
\addplot[scatter, scatter src = explicit symbolic, only marks, mark size = 1, opacity = 0.5, mark = *, mark options = {line width = 0pt}] table[x index = 0, y index = 1, meta index = 2, header = false]
{alon_0.26_0.25_supmean.sct};
\pgfplotsextra{\yafdrawaxis{\dxmin}{\dxmax}{\dymin}{\dymax}}
\end{axis}
\begin{axis}[at = {(scatter.north west)},
    width = \imgwidth,
    height = 0.5cm,
	scale only axis,
	xmin = \dxmin, xmax = \dxmax,
	ytick = \empty, xtick = \empty, scaled ticks = false,
    ]
\addplot+[const plot, no markers, yafcolor5, fill = yafcolor5!50] table[x index = 1, y index = 0, header = false] {alon_0.26_0.25_supmean.hst} \closedcycle;
\end{axis}
\begin{axis}[at = {(scatter.south east)},
    height = \imgheight,
    width = 0.5cm,
	scale only axis,
	ymin = \dymin, ymax = \dymax,
	ytick = \empty, xtick = \empty, scaled ticks = false,
    ]
\addplot+[const plot, no markers, yafcolor5, fill = yafcolor5!50] table[x index = 2, y index = 3, header = false] {alon_0.26_0.25_supmean.hst} \closedcycley;
\end{axis}

\end{tikzpicture} &
\begin{tikzpicture}
\pgfmathsetmacro{\imgwidth}{\imgsize}
\pgfmathsetmacro{\imgheight}{\imgsize}
\pgfmathsetmacro{\dxmin}{0.03}
\pgfmathsetmacro{\dxmax}{0.3}
\pgfmathsetmacro{\dymin}{0.08}
\pgfmathsetmacro{\dymax}{0.367}

\begin{axis}[xlabel={$\cp{X; D}$}, title = {\scriptsize(d) \emph{Thalia}}, title style = {anchor = south east, at = {(1, 1)}}, 
	name = scatter,
    width = \imgwidth,
    height = \imgheight,
	scale only axis,
    cycle list name=yaf,
	scatter/classes = {2={yafcolor1},3={yafcolor2},4={yafcolor3},5={yafcolor4},6={yafcolor7}},
	xmin = \dxmin, xmax = \dxmax, ymin = \dymin, ymax = \dymax,
    ]
\addplot[scatter, scatter src = explicit symbolic, only marks, mark size = 1, opacity = 0.5, mark = *, mark options = {line width = 0pt}] table[x index = 0, y index = 1, meta index = 2, header = false]
{thalia_0.08_0.25_supmean.sct};
\pgfplotsextra{\yafdrawaxis{\dxmin}{\dxmax}{\dymin}{\dymax}}
\end{axis}
\begin{axis}[at = {(scatter.north west)},
    width = \imgwidth,
    height = 0.5cm,
	scale only axis,
	xmin = \dxmin, xmax = \dxmax,
	ytick = \empty, xtick = \empty, scaled ticks = false,
    ]
\addplot+[const plot, no markers, yafcolor5, fill = yafcolor5!50] table[x index = 1, y index = 0, header = false] {thalia_0.08_0.25_supmean.hst} \closedcycle;
\end{axis}
\begin{axis}[at = {(scatter.south east)},
    height = \imgheight,
    width = 0.5cm,
	scale only axis,
	ymin = \dymin, ymax = \dymax,
	ytick = \empty, xtick = \empty, scaled ticks = false,
    ]
\addplot+[const plot, no markers, yafcolor5, fill = yafcolor5!50] table[x index = 2, y index = 3, header = false] {thalia_0.08_0.25_supmean.hst} \closedcycley;
\end{axis}

\end{tikzpicture} &
\begin{tikzpicture}
\pgfmathsetmacro{\imgwidth}{\imgsize}
\pgfmathsetmacro{\imgheight}{\imgsize}
\pgfmathsetmacro{\dxmin}{0.08}
\pgfmathsetmacro{\dxmax}{0.33}
\pgfmathsetmacro{\dymin}{0.2}
\pgfmathsetmacro{\dymax}{0.41}

\begin{axis}[xlabel={$\cp{X; D}$}, title = {\scriptsize(e) \emph{Yeast}}, title style = {anchor = south east, at = {(1, 1)}}, 
	name = scatter,
    width = \imgwidth,
    height = \imgheight,
	scale only axis,
    cycle list name=yaf,
	scatter/classes = {2={yafcolor1},3={yafcolor2},4={yafcolor3},5={yafcolor4},6={yafcolor7},7={yafcolor5},8={yafcolor8}},
	xmin = \dxmin, xmax = \dxmax, ymin = \dymin, ymax = \dymax,
    ]
\addplot[scatter, scatter src = explicit symbolic, only marks, mark size = 1, opacity = 0.5, mark = *, mark options = {line width = 0pt}] table[x index = 0, y index = 1, meta index = 2, header = false]
{yeast_0.20_0.25_supmean.sct};
\pgfplotsextra{\yafdrawaxis{\dxmin}{\dxmax}{\dymin}{\dymax}}
\end{axis}
\begin{axis}[at = {(scatter.north west)},
    width = \imgwidth,
    height = 0.5cm,
	scale only axis,
	xmin = \dxmin, xmax = \dxmax,
	ytick = \empty, xtick = \empty, scaled ticks = false,
    ]
\addplot+[const plot, no markers, yafcolor5, fill = yafcolor5!50] table[x index = 1, y index = 0, header = false] {yeast_0.20_0.25_supmean.hst} \closedcycle;
\end{axis}
\begin{axis}[at = {(scatter.south east)},
    height = \imgheight,
    width = 0.5cm,
	scale only axis,
	ymin = \dymin, ymax = \dymax,
	ytick = \empty, xtick = \empty, scaled ticks = false,
    ]
\addplot+[const plot, no markers, yafcolor5, fill = yafcolor5!50] table[x index = 2, y index = 3, header = false] {yeast_0.20_0.25_supmean.hst} \closedcycley;
\end{axis}

\end{tikzpicture} \\
\begin{tikzpicture}
\pgfmathsetmacro{\imgwidth}{\imgsize}
\pgfmathsetmacro{\imgheight}{\imgsize}
\pgfmathsetmacro{\dxmin}{-5}
\pgfmathsetmacro{\dxmax}{5}
\pgfmathsetmacro{\dymin}{-5}
\pgfmathsetmacro{\dymax}{5}

\begin{axis}[xlabel={$\zind{X; D}$}, ylabel= {$\zpart{X; D}$},
	name = scatter,
    width = \imgwidth,
    height = \imgheight,
	scale only axis,
    cycle list name=yaf,
	scatter/classes = {2={yafcolor1},3={yafcolor2},4={yafcolor3},5={yafcolor4},6={yafcolor7}},
	xmin = \dxmin, xmax = \dxmax, ymin = \dymin, ymax = \dymax,
    ]
\addplot[scatter, scatter src = explicit symbolic, only marks, mark size = 1, opacity = 0.5, mark = *, mark options = {line width = 0pt}] table[x index = 0, y index = 1, meta index = 2, header = false]
	{ind_0.1_0.25_indpart.sct};
\pgfplotsextra{\yafdrawaxis{\dxmin}{\dxmax}{\dymin}{\dymax}}
\end{axis}
\begin{axis}[at = {(scatter.north west)},
    width = \imgwidth,
    height = 0.5cm,
	scale only axis,
	xmin = \dxmin, xmax = \dxmax,
	ytick = \empty, xtick = \empty, scaled ticks = false,
    ]
\addplot+[const plot, no markers, yafcolor5, fill = yafcolor5!50] table[x index = 1, y index = 0, header = false]
	{ind_0.1_0.25_indpart.hst} \closedcycle;
\end{axis}
\begin{axis}[at = {(scatter.south east)},
    height = \imgheight,
    width = 0.5cm,
	scale only axis,
	ymin = \dymin, ymax = \dymax,
	ytick = \empty, xtick = \empty, scaled ticks = false,
    ]
\addplot+[const plot, no markers, yafcolor5, fill = yafcolor5!50] table[x index = 2, y index = 3, header = false]
	{ind_0.1_0.25_indpart.hst} \closedcycley;
\end{axis}
\end{tikzpicture} &
\begin{tikzpicture}
\pgfmathsetmacro{\imgwidth}{\imgsize}
\pgfmathsetmacro{\imgheight}{\imgsize}
\pgfmathsetmacro{\dxmin}{-5}
\pgfmathsetmacro{\dxmax}{105}
\pgfmathsetmacro{\dymin}{-5}
\pgfmathsetmacro{\dymax}{45}

\tikzset{every pin/.style={circle,font=\scriptsize, color = black, pin distance = 4pt, inner sep = 0pt, outer sep = 0pt, pin edge = {thick}}}

\begin{axis}[xlabel={$\zind{X; D}$},
	name = scatter,
    width = \imgwidth,
    height = \imgheight,
	scale only axis,
    cycle list name=yaf,
	scatter/classes = {2={yafcolor1},3={yafcolor2},4={yafcolor3},5={yafcolor4},6={yafcolor7}},
	xmin = \dxmin, xmax = \dxmax, ymin = \dymin, ymax = \dymax,
    ]
\addplot[scatter, scatter src = explicit symbolic, only marks, mark size = 1, opacity = 0.5, mark = *, mark options = {line width = 0pt}] table[x index = 0, y index = 1, meta index = 2, header = false]
	{plant_0.08_0.25_indpart.sct};
\node[circle, inner sep = 1pt, pin={110:$\mathcal{A}$}] at (axis cs: 100.43,  12.711) {};
\node[circle, inner sep = 1pt, pin={0:$\mathcal{B}$}] at (axis cs: 75.20, 0.31) {};
\node[circle, inner sep = 1pt, pin={270:$\mathcal{C}$}] at (axis cs: 38, 38) {};
\node[circle, inner sep = 1pt, pin={90:$\mathcal{D}$}] at (axis cs: 71.03, 16.08) {};
\pgfplotsextra{\yafdrawaxis{\dxmin}{\dxmax}{\dymin}{\dymax}}
\end{axis}
\begin{axis}[at = {(scatter.north west)},
    width = \imgwidth,
    height = 0.5cm,
	scale only axis,
	xmin = \dxmin, xmax = \dxmax,
	ytick = \empty, xtick = \empty, scaled ticks = false,
    ]
\addplot+[const plot, no markers, yafcolor5, fill = yafcolor5!50] table[x index = 1, y index = 0, header = false]
	{plant_0.08_0.25_indpart.hst} \closedcycle;
\end{axis}
\begin{axis}[at = {(scatter.south east)},
    height = \imgheight,
    width = 0.5cm,
	scale only axis,
	ymin = \dymin, ymax = \dymax,
	ytick = \empty, xtick = \empty, scaled ticks = false,
    ]
\addplot+[const plot, no markers, yafcolor5, fill = yafcolor5!50] table[x index = 2, y index = 3, header = false]
	{plant_0.08_0.25_indpart.hst} \closedcycley;
\end{axis}
\end{tikzpicture} &
\begin{tikzpicture}
\pgfmathsetmacro{\imgwidth}{\imgsize}
\pgfmathsetmacro{\imgheight}{\imgsize}
\pgfmathsetmacro{\dxmin}{15}
\pgfmathsetmacro{\dxmax}{200}
\pgfmathsetmacro{\dymin}{6}
\pgfmathsetmacro{\dymax}{42}

\begin{axis}[xlabel={$\zind{X; D}$},
	name = scatter,
    width = \imgwidth,
    height = \imgheight,
	scale only axis,
    cycle list name=yaf,
	scatter/classes = {2={yafcolor1},3={yafcolor2},4={yafcolor3},5={yafcolor4},6={yafcolor7}},
	xmin = \dxmin, xmax = \dxmax, ymin = \dymin, ymax = \dymax,
    ]
\addplot[scatter, scatter src = explicit symbolic, only marks, mark size = 1, opacity = 0.5, mark = *, mark options = {line width = 0pt}] table[x index = 0, y index = 1, meta index = 2, header = false]
	{alon_0.26_0.25_indpart.sct};
\pgfplotsextra{\yafdrawaxis{\dxmin}{\dxmax}{\dymin}{\dymax}}
\end{axis}
\begin{axis}[at = {(scatter.north west)},
    width = \imgwidth,
    height = 0.5cm,
	scale only axis,
	xmin = \dxmin, xmax = \dxmax,
	ytick = \empty, xtick = \empty, scaled ticks = false,
    ]
\addplot+[const plot, no markers, yafcolor5, fill = yafcolor5!50] table[x index = 1, y index = 0, header = false]
	{alon_0.26_0.25_indpart.hst} \closedcycle;
\end{axis}
\begin{axis}[at = {(scatter.south east)},
    height = \imgheight,
    width = 0.5cm,
	scale only axis,
	ymin = \dymin, ymax = \dymax,
	ytick = \empty, xtick = \empty, scaled ticks = false,
    ]
\addplot+[const plot, no markers, yafcolor5, fill = yafcolor5!50] table[x index = 2, y index = 3, header = false]
	{alon_0.26_0.25_indpart.hst} \closedcycley;
\end{axis}
\end{tikzpicture} &
\begin{tikzpicture}
\pgfmathsetmacro{\imgwidth}{\imgsize}
\pgfmathsetmacro{\imgheight}{\imgsize}
\pgfmathsetmacro{\dxmin}{-5}
\pgfmathsetmacro{\dxmax}{32}
\pgfmathsetmacro{\dymin}{-5}
\pgfmathsetmacro{\dymax}{16}

\begin{axis}[xlabel={$\zind{X; D}$},
	name = scatter,
    width = \imgwidth,
    height = \imgheight,
	scale only axis,
    cycle list name=yaf,
	scatter/classes = {2={yafcolor1},3={yafcolor2},4={yafcolor3},5={yafcolor4},6={yafcolor7}},
	xmin = \dxmin, xmax = \dxmax, ymin = \dymin, ymax = \dymax,
    ]
\addplot[scatter, scatter src = explicit symbolic, only marks, mark size = 1, opacity = 0.5, mark = *, mark options = {line width = 0pt}] table[x index = 0, y index = 1, meta index = 2, header = false]
	{thalia_0.12_0.25_indpart.sct};
\pgfplotsextra{\yafdrawaxis{\dxmin}{\dxmax}{\dymin}{\dymax}}
\end{axis}
\begin{axis}[at = {(scatter.north west)},
    width = \imgwidth,
    height = 0.5cm,
	scale only axis,
	xmin = \dxmin, xmax = \dxmax,
	ytick = \empty, xtick = \empty, scaled ticks = false,
    ]
\addplot+[const plot, no markers, yafcolor5, fill = yafcolor5!50] table[x index = 1, y index = 0, header = false]
	{thalia_0.12_0.25_indpart.hst} \closedcycle;
\end{axis}
\begin{axis}[at = {(scatter.south east)},
    height = \imgheight,
    width = 0.5cm,
	scale only axis,
	ymin = \dymin, ymax = \dymax,
	ytick = \empty, xtick = \empty, scaled ticks = false,
    ]
\addplot+[const plot, no markers, yafcolor5, fill = yafcolor5!50] table[x index = 2, y index = 3, header = false]
	{thalia_0.12_0.25_indpart.hst} \closedcycley;
\end{axis}
\end{tikzpicture} &
\begin{tikzpicture}
\pgfmathsetmacro{\imgwidth}{\imgsize}
\pgfmathsetmacro{\imgheight}{\imgsize}
\pgfmathsetmacro{\dxmin}{-22}
\pgfmathsetmacro{\dxmax}{480}
\pgfmathsetmacro{\dymin}{-22}
\pgfmathsetmacro{\dymax}{55}

\begin{axis}[xlabel={$\zind{X; D}$},
	name = scatter,
    width = \imgwidth,
    height = \imgheight,
	scale only axis,
    cycle list name=yaf,
	scatter/classes = {2={yafcolor1},3={yafcolor2},4={yafcolor3},5={yafcolor4},6={yafcolor7},7={yafcolor5},8={yafcolor8}},
	xmin = \dxmin, xmax = \dxmax, ymin = \dymin, ymax = \dymax,
    ]
\addplot[scatter, scatter src = explicit symbolic, only marks, mark size = 1, opacity = 0.5, mark = *, mark options = {line width = 0pt}] table[x index = 0, y index = 1, meta index = 2, header = false]
	{yeast_0.20_0.25_indpart.sct};
\pgfplotsextra{\yafdrawaxis{\dxmin}{\dxmax}{\dymin}{\dymax}}
\end{axis}
\begin{axis}[at = {(scatter.north west)},
    width = \imgwidth,
    height = 0.5cm,
	scale only axis,
	xmin = \dxmin, xmax = \dxmax,
	ytick = \empty, xtick = \empty, scaled ticks = false,
    ]
\addplot+[const plot, no markers, yafcolor5, fill = yafcolor5!50] table[x index = 1, y index = 0, header = false]
	{yeast_0.20_0.25_indpart.hst} \closedcycle;
\end{axis}
\begin{axis}[at = {(scatter.south east)},
    height = \imgheight,
    width = 0.5cm,
	scale only axis,
	ymin = \dymin, ymax = \dymax,
	ytick = \empty, xtick = \empty, scaled ticks = false,
    ]
\addplot+[const plot, no markers, yafcolor5, fill = yafcolor5!50] table[x index = 2, y index = 3, header = false]
	{yeast_0.20_0.25_indpart.hst} \closedcycley;
\end{axis}
\end{tikzpicture} &
\begin{tikzpicture}[baseline = -2.7cm]
\node[inner sep = 0.5pt, text width = 20pt, align = right] (n0) {\scriptsize pattern\\size\\};
\node[circle, line width = 0.5pt, inner sep = 0.5pt, text width = 3pt, draw = yafcolor1, fill = yafcolor1!50, label = {right: \scriptsize 2},  below = 4pt of n0] (n1) {};
\node[circle, line width = 0.5pt, inner sep = 0.5pt, text width = 3pt, draw = yafcolor2, fill = yafcolor2!50, label = {right: \scriptsize 3},  below = 2pt of n1] (n2) {};
\node[circle, line width = 0.5pt, inner sep = 0.5pt, text width = 3pt, draw = yafcolor3, fill = yafcolor3!50, label = {right: \scriptsize 4},  below = 2pt of n2] (n3) {};
\node[circle, line width = 0.5pt, inner sep = 0.5pt, text width = 3pt, draw = yafcolor4, fill = yafcolor4!50, label = {right: \scriptsize 5},  below = 2pt of n3] (n4) {};
\node[circle, line width = 0.5pt, inner sep = 0.5pt, text width = 3pt, draw = yafcolor7, fill = yafcolor7!50, label = {right: \scriptsize 6},  below = 2pt of n4] (n5) {};
\node[circle, line width = 0.5pt, inner sep = 0.5pt, text width = 3pt, draw = yafcolor5, fill = yafcolor5!50, label = {right: \scriptsize 7},  below = 2pt of n5] (n6) {};
\node[circle, line width = 0.5pt, inner sep = 0.5pt, text width = 3pt, draw = yafcolor8, fill = yafcolor8!50, label = {right: \scriptsize 8},  below = 2pt of n6] (n7) {};
\end{tikzpicture}
\\
\end{tabular}
\caption{Scatter plots and histograms of supports and ranks. Each plot contains
a scatter plot of two variables and the corresponding marginal histograms. The
top row contains $\cp{X; D, 0.25}$ plotted as a function of $\cp{X; D}$. The bottom row
contains contains $\zpart{X; D}$ as a function of $\zind{X; D}$. Each column corresponds
to a single dataset. Pattern sizes are encoded with different colours.
Note that the axis' ranges vary.}
\label{fig:scatter}
\end{figure*}

\emph{Setup:} For each dataset we computed frequent itemsets using $\cp{X; D,
0.25}$ as a support. We set the threshold such that we get roughly several
hundred thousand itemsets, see Table~\ref{tab:basic}.
We then ranked itemsets using $\zind{X}$ and $\zpart{X}$. The results are given
in Figure~\ref{fig:scatter}.

\emph{Support comparison:}
Let us first compare supports $\cp{X}$ and $\cp{X; 0.25}$, given in the top row
of Figure~\ref{fig:scatter}. We see that for a fixed itemset length
there is a strong linear correlation between the supports.
The histograms reveal why we should consider $\cp{X;
0.25}$ as a stopping criterion instead of $\cp{X}$. A significantly large
number of itemsets of length, say $M$, will have larger support than any
itemset of length $M + 1$ or higher, that is, in order to discover any itemset
of length $3$, we will have to discover all itemsets of length $2$. This problem
does not occur with $\cp{X; D, 0.25}$. 

\emph{Normalisation comparison:}
Our next step is to compare ranks, given on the second row of
Figure~\ref{fig:scatter}. As expected $\zpart{X}$ is more conservative than
$\zind{X}$. For example, in \emph{Ind}, $\zind{X}$ is distributed as $N(0, 1)$,
as predicted by Proposition~\ref{prop:normality}, whereas $\zpart{X}$ is skewed
towards negative values. In general, $\zind{X}$ prefers large itemsets whereas
$\zpart{X}$ prefers small ones. This can be beneficial as seen with
\emph{Plant} dataset. The first $5$ itemsets $\mathcal{A}$ according to
$\zind{X}$ are the itemsets related to subspace clusters.  However, the next
itemsets $\mathcal{B}$ are the clusters with some additional unrelated items,
on the other hand, $\zpart{X}$ will assign a low score to $\mathcal{B}$. In
addition, $\zpart{X}$ favours sets $\mathcal{C}$ and $\mathcal{D}$ itemsets of
size $2$--$3$ that are subitemsets of $\mathcal{A}$.

\emph{Computational complexity:}
While optimising for speed is not the focus in this work,
our implementation\footnote{Python implementation available at \url{http://users.ics.aalto.fi/ntatti/}} is able to discover
several hundred thousand patterns in minutes. The datasets we consider here are
relatively small when compared to the size of the binary datasets used for
mining normal patterns. However, the speed of traditional miners is based on
the fact that binary datasets are typically very sparse. We do not have the
same luxury and computing of each itemset requires a full scan. On the
other hand, the cost for computing the support a single itemset
depends only on the size of the itemset and the number of datapoints whereas
the performance of traditional itemset miners depends heavily on how 1s are
distributed in the dataset.

\section{Concluding Remarks}\label{sec:conclusions}
In this paper we proposed a measure of quality for itemsets mined from
real-valued dataset. Our approach was to compute the average support from
binarized data with random thresholds. Despite the complex definition we can
compute the measure efficiently. As a distribution for a threshold we
considered a special distribution related to copulas. We normalised the support
by comparing the observed support to the expected support according to a null
hypothesis. We considered two hypotheses: the first assumption is that all
items are independent, while the second assumption is more general---we
assume that items are independent w.r.t. to a given partition.

This research opens up several directions for future work.

Firstly, we
considered one specific threshold distribution. This distribution is a good
choice if you do not have any information about the distribution of individual
items. However, there are other choices. For example, if we know that data is
distributed between $[a, b]$, we can consider a uniform distribution over the
interval, see~\cite{DBLP:conf/sdm/JaroszewiczK07}, or possibly a shorter interval that excludes the extreme values.

The speed-up techniques used for mining sparse binary data no longer apply. 
This raises a question whether we can speed up significantly the mining procedure.

Lastly, the distribution of itemsets is different than of those that are
obtained from binary data. Typically, in binary data, the margins of the items
are distributed unevenly: there will be a lot of items that are rare and some
items that are frequent. This means a lot of itemsets will be pruned in first
steps.  This is not the case with the copula support, where typically you will pass
almost all items of size $2$. This emphasizes the need for ranking itemsets, in
our case, we used $\zind{X}$ and $\zpart{X}$. However, as future work it
would be interesting to see what type of constraints one can impose on itemsets
in order to reduce the output.

\section*{Acknowledgment}
This work was supported by Academy of Finland grant 118653 ({\sc algodan})

\bibliographystyle{IEEEtran}
\bibliography{bibliography}

\begin{thebibliography}{10}
\providecommand{\url}[1]{#1}
\csname url@samestyle\endcsname
\providecommand{\newblock}{\relax}
\providecommand{\bibinfo}[2]{#2}
\providecommand{\BIBentrySTDinterwordspacing}{\spaceskip=0pt\relax}
\providecommand{\BIBentryALTinterwordstretchfactor}{4}
\providecommand{\BIBentryALTinterwordspacing}{\spaceskip=\fontdimen2\font plus
\BIBentryALTinterwordstretchfactor\fontdimen3\font minus
  \fontdimen4\font\relax}
\providecommand{\BIBforeignlanguage}[2]{{%
\expandafter\ifx\csname l@#1\endcsname\relax
\typeout{** WARNING: IEEEtran.bst: No hyphenation pattern has been}%
\typeout{** loaded for the language `#1'. Using the pattern for}%
\typeout{** the default language instead.}%
\else
\language=\csname l@#1\endcsname
\fi
#2}}
\providecommand{\BIBdecl}{\relax}
\BIBdecl

\bibitem{nelsen:06:copula}
R.~B. Nelsen, \emph{An introduction to Copulas}.\hskip 1em plus 0.5em minus
  0.4em\relax Springer, 2006.

\bibitem{agrawal:96:fast}
R.~Agrawal, H.~Mannila, R.~Srikant, H.~Toivonen, and A.~I. Verkamo, ``Fast
  discovery of association rules,'' in \emph{Advances in Knowledge Discovery
  and Data Mining}, 1996, pp. 307--328.

\bibitem{DBLP:journals/tkdd/Webb10}
G.~I. Webb, ``Self-sufficient itemsets: An approach to screening potentially
  interesting associations between items,'' \emph{TKDD}, vol.~4, no.~1, pp.
  3:1--3:20, 2010.

\bibitem{DBLP:conf/sigmod/SrikantA96}
R.~Srikant and R.~Agrawal, ``Mining quantitative association rules in large
  relational tables,'' in \emph{SIGMOD}, 1996, pp. 1--12.

\bibitem{DBLP:conf/kdd/CaldersGJ06}
T.~Calders, B.~Goethals, and S.~Jaroszewicz, ``Mining rank-correlated sets of
  numerical attributes,'' in \emph{KDD}, 2006, pp. 96--105.

\bibitem{DBLP:conf/sdm/JaroszewiczK07}
S.~Jaroszewicz and M.~Korzen, ``Approximating representations for large
  numerical databases,'' in \emph{SDM}, 2007.

\bibitem{DBLP:conf/kdd/SteinbachTXK04}
M.~Steinbach, P.-N. Tan, H.~Xiong, and V.~Kumar, ``Generalizing the notion of
  support,'' in \emph{KDD}, 2004, pp. 689--694.

\bibitem{brin:97:beyond}
S.~Brin, R.~Motwani, and C.~Silverstein, ``Beyond market baskets: Generalizing
  association rules to correlations,'' in \emph{SIGMOD}, 1997, pp. 265--276.

\bibitem{jaroszewicz:04:interestingness}
S.~Jaroszewicz and D.~A. Simovici, ``Interestingness of frequent itemsets using
  bayesian networks as background knowledge,'' in \emph{KDD}, 2004, pp.
  178--186.

\bibitem{tatti:08:maximum}
N.~Tatti, ``Maximum entropy based significance of itemsets,'' \emph{KAIS},
  vol.~17, no.~1, pp. 57--77, 2008.

\bibitem{DBLP:conf/pakdd/ChuiKH07}
C.~K. Chui, B.~Kao, and E.~Hung, ``Mining frequent itemsets from uncertain
  data,'' in \emph{PAKDD}, 2007, pp. 47--58.

\bibitem{DBLP:conf/kdd/BerneckerKRVZ09}
T.~Bernecker, H.-P. Kriegel, M.~Renz, F.~Verhein, and A.~Z{\"u}fle,
  ``Probabilistic frequent itemset mining in uncertain databases,'' in
  \emph{KDD}, 2009, pp. 119--128.

\bibitem{DBLP:conf/icdm/CaldersGG10}
T.~Calders, C.~Garboni, and B.~Goethals, ``Approximation of frequentness
  probability of itemsets in uncertain data,'' in \emph{ICDM}, 2010, pp.
  749--754.

\bibitem{reshef:01:novel}
D.~N. Reshef, Y.~A. Reshef, H.~K. Finucane, S.~R. Grossman, G.~McVean, P.~J.
  Turnbaugh, E.~S. Lander, M.~Mitzenmacher, and P.~C. Sabeti, ``Detecting novel
  associations in large data sets,'' \emph{Science}, vol. 334, no. 6062, pp.
  518--1524, 2011.

\bibitem{szekely:09:distance}
G.~J. Sz\'{e}kely and M.~L. Rizzo, ``Brownian distance covariance,'' \emph{The
  Annals of Applied Statistics}, vol.~3, no.~4, pp. 1236--1265, 2009.

\bibitem{alon:99:pnas}
U.~Alon, N.~Barkai, D.~A. Notterman, K.~Gish, D.~M. S.~Ybarra, and A.~J.
  Levine, ``Broad patterns of gene expression revealed by clustering of tumor
  and normal colon tissues probed by oligonucleotide arrays,'' \emph{PNAS},
  vol.~96, no.~12, pp. 6745--6750, 1999.

\bibitem{vaart:98:asymptotic}
A.~W. van~der Vaart, \emph{Asymptotic Statistics}.\hskip 1em plus 0.5em minus
  0.4em\relax Cambridge University Press, 1998.

\end{thebibliography}

\appendix
\label{sec:appendix}
\subsection{Proof of Proposition~\ref{prop:normality}}

In order to prove the proposition we need the following proposition.

\begin{proposition}[Theorem~12.3~in~\cite{vaart:98:asymptotic}]
\label{prop:ustat} Let $h$ be a function (called kernel) of $L$ parameters. Assume that $h$
is symmetric w.r.t. its parameters (that is, any permutation of parameters will yield the same result).
Let $Y_1, \ldots, Y_N$ be $N$ i.i.d. variables such that $\mean{h^2(Y_1, \ldots, Y_L)} < \infty$. Then
\[
	\frac{\sqrt{N}}{{N \choose L}} \sum_{i_1, \ldots, i_L} h(Y_{i_1}, \ldots, Y_{i_L}) - \mu,
\]
where the sum goes over all subsets of size $L$ and $\mu = \mean{h(Y_1, \ldots, Y_L)} $, approaches a Gaussian
distribution as $N$ goes to infinity.
\end{proposition}

\begin{IEEEproof}[Proof of Proposition~\ref{prop:normality}]
Note that since $S_{ij}$ and $S_{kj}$ are not independent, we cannot use
Central Limit theorem to prove normality.
Instead we will use $U$-statistics to prove the result. In order to do that let
us first define a function of $M + 1$ vectors of length $M$,
\[
	g(y_0, y_1, \ldots, y_M) = \prod_{i = 1}^M \indi{y_{0x_i} > y_{ix_i}},
\]
where $x_i$ are the items of $X = x_1\cdots x_M$.
Note that
\[
	U = \frac{1}{N(N - 1)^M}\sum_{i_0 = 1}^N \cdots \sum_{i_M = 1, \atop i_M \neq i_0}^N g(Y_{i_0}, \ldots, Y_{i_M})\quad.
\]

Proposition~\ref{prop:ustat} requires a kernel to be symmetric w.r.t. its
parameters. In order to do that, let us define 
\[
	h(y_0, y_1, \ldots, y_M) = \sum_{\tau} g(y_{\tau(0)}, \ldots, y_{\tau(M)}),
\]
where the sum goes over all permutations $\tau$ of size $M + 1$.
Then according to Proposition~\ref{prop:ustat} a statistic $\sqrt{N}U'$, where
\[
	U' = \frac{1}{{N \choose M + 1}}\sum_{i_0, \ldots, i_M} h(Y_{i_0}, \ldots, Y_{i_M}),
\]
where the sum goes over all $M + 1$ subsets of $\enpr{1}{N}$,
converges to a Gaussian distribution.

The statistics $U$ and $U'$ have the same mean, say $\mu = \mean{U} = \mean{U'}$, but they are different. We will show next that this
difference becomes minute as $N$ approaches infinity.
In order to do that, let us define 
\[
	a(N) = N(N - 1)^M \quad\text{and}\quad b(N) = {N \choose M + 1}(M + 1)!\quad.
\]

The sum of $U'$ requires
that all rows $Y_{i_k}$ for must be different where as $U$ only
requires that $Y_{i_0}$ is different from the remaining rows. Hence, there are
$a(N) - b(N)$ less terms in $U'$. Let $Z$ be the sum of these terms.
We have
\[
	U  = \frac{b(N)}{a(N)}U' +  \frac{Z}{a(N)} \quad.
\]
Let us write $r(N) = (a(N) - b(N)) / a(N)$.
Both $a(N)$ and $b(N)$ are polynomials of degree $M + 1$ 
and the coefficient of the highest term is $1$ for both polynomials.
Consequently, $a(N) - b(N)$ is a polynomial of degree $M$.
This implies that $r(N)$ and $r(N)\sqrt{N}$ both go to $0$ as $N$ approaches infinity.

We can express the difference as
\[
	\sqrt{N}(U - U') = r(N)\sqrt{N}(U' - \mu) + r(N)\sqrt{N}\mu  + \frac{Z}{a(N)},
\]
According to Proposition~\ref{prop:ustat}, $\sqrt{N}(U' - \mu)$
converges to a Gaussian distribution and since $r(N)$ converges to $0$,
it follows that the first term goes to $0$ as $N$ goes to infinity.
Similarly, the second term goes to $0$ since $r(N)\sqrt{N}$ goes to $0$.
Finally, to bound the last term note that
\[
	0 \leq \frac{Z}{a(N)} \leq \frac{a(N) - b(N)}{a(N)} = r(N)
\]
which implies that $Z / a(N)$ goes to $0$ as $N$ approaches infinity.
We have shown that $\sqrt{N}(U - \theta)$ and $\sqrt{N}(U' - \theta)$
converge to each other in probability and that the latter approaches a Gaussian distribution.
\end{IEEEproof}

\subsection{Proof of Proposition~\ref{prop:partvar}}
First, we need the following technical proposition.

\begin{proposition}
\label{prop:decompose}
Assume that we a given integers $N$ and $K$ let $\Omega = [1, \ldots, N]^K$
be the set of integer vectors of length $K$. Let $\funcdef{f_N}{\Omega}{[0, 1]}$ be a
function  such that $\max_{\omega \in \Omega} \abs{f_N(\omega)} \in O(N^{-K + 1})$.
Let $P \subset \Omega$ be the subset containing only vectors with distinct entries.
Assume that we are given $K(K - 1)/2$ subsets
$\Omega_{ij}$ such that
\[
	\set{\omega \in \Omega \mid \omega_i = \omega_j} \subseteq \Omega_{ij} \subseteq \Omega \setminus P,
\]
that is, $\Omega_{ij}$ contains vectors for which $i$th and $j$th entries are the same and has no vectors from $P$.
Then
\[
	\sum_{\omega \in \Omega \setminus P} f(\omega) \to \sum_{i = 1}^{K - 1} \sum_{j = i + 1}^K \sum_{\omega \in \Omega_{ij}} f(\omega)
\]
as $N$ approaches infinity.
\end{proposition}

\begin{IEEEproof}
Partition $\Omega$ into $K$ groups $\Omega^0, \ldots, \Omega^{K - 1}$ such that
\[
	\Omega^i = \set{\omega \in \Omega \mid \omega \text{ has } K - i \text{ distinct entries }}\quad.
\]
A direct computation shows that
\[
	\abs{\Omega^i} = S(K, K - i) \frac{N!}{(N - K + i)!} \in O(N^{K - i}),
\]
where $S(K, K - i)$ is a Sterling number of the second kind.
Let $\Theta = \Omega^2 \cup \cdots \cup \Omega^{K - 1}$.
This immediately implies that $\abs{\Theta} \in O(N^{K - 2})$.
Note that $P = \Omega^0$ and that $\Omega^1 \subseteq \bigcup_{i,j} \Omega_{ij}$.
Let $\Delta_{ij} = \Omega_{ij} \setminus \Omega_1$ be the set of vectors that have $i$ and $j$
as common entries and have less than $K - 1$ unique entries. Note that $\Delta_{ij} \subset \Theta$.
Consequently $\abs{\Delta_{ij}} \in O(N^{K - 2})$. We can now write the sum as
\[
	\sum_{\omega \in \Omega \setminus P} f(\omega) =
	\sum_{i \neq j} \sum_{\omega \in \Omega_{ij}} f(\omega) 
	- \sum_{i \neq j} \sum_{\omega \in \Delta_{ij}} f(\omega) 
	+ \sum_{\omega \in \Theta} f(\omega)\ .
\]
Since $f$ is bounded by $O(N^{- K + 1})$, 
the third and fourth terms vanish as $N$ goes to infinity. 
\end{IEEEproof}

\begin{IEEEproof}[Proof of Proposition~\ref{prop:partvar}]
For notational simplicity, let us assume that
$X = 1\cdots M$ and define $f$ and $g$ as functions of $2M + 2$ variables,
\[
	f(i_0, \ldots, i_M, j_0, \ldots, j_M) = \frac{1}{N}\meanb{\prod_{k = 1}^M O_{i_0i_kk}\prod_{k = 1}^M O_{j_0j_kk}  }
\]
and
\[
	g(i_0, \ldots, j_0, \ldots, j_M) = \frac{1}{N}\meanb{\prod_{k = 1}^M O_{i_0i_kk}}\meanb{\prod_{k = 1}^M O_{j_0j_kk}  }\quad.
\]
Let $\Omega = [1, \ldots, N]^{2M + 2}$. Note that
\[
	\meanb{NU^2} = \sum_{\omega \in \Omega} f(\omega)
\quad\text{and}\quad
	\meanb{\sqrt{N}U}^2 = \sum_{\omega \in \Omega} g(\omega)\quad.
\]
Let $\Omega^i$ as defined in Proposition~\ref{prop:decompose}.
Let us define
\[
	\Omega_{ij} = \set{\omega \in \Omega \mid \omega_i = \omega_j}
\]
for $1 \leq i \leq M$ and $M + 1 \leq j \leq 2M$, and also
\[
	\Omega_{ij} = \set{\omega \in \Omega^1 \mid \omega_i = \omega_j}
\]
whenever $1 \leq i, j \leq M$ or $M + 1 \leq i, j \leq 2M$. 
Note that $f(\omega) = g(\omega)$ whenever $\omega$ does not share
any entry between the first $M$ entries and the last $M$ entries.
This holds when $\omega \in \Omega^0$ or when $\omega \in \Omega_{ij}$
for $1 \leq i, j \leq M$ or $M + 1 \leq i, j \leq 2M$.

We can now apply Proposition~\ref{prop:decompose} to conclude
that
\[
	\sum_{\omega \in \Omega} f(\omega) - \sum_{\omega \in \Omega} g(\omega) \to \sum_{i = 1}^M \sum_{j = M + 1}^{2M} \sum_{\omega \in \Omega_{ij}} \pr{f(\omega) - g(\omega)}
\]
as $N$ approaches infinity.

Our final step is to compute the sums in the above equation.

First note that
\[
	 \sum_{\omega \in \Omega_{1(M + 1)}} f(\omega) - g(\omega) = \frac{1}{N}\sum_{k = 1}^N \meanb{T_k^2} - \meanb{T_k}^2 = \alpha - \mu^2\ .
\]

Let $1 < i, j \leq M$.  Then
\[
\begin{split}
	 \sum_{\omega \in \Omega_{i(j + M)}} \!\!\!\!f(\omega) & = \frac{1}{N}\sum_{k, l, m = 1}^N \mean{T_{ki}T_{lj}O_{kmi}O_{lmj}} \\
	 & = \frac{1}{N}\sum_{m = 1}^N \meanb{\big(\sum_{k = 1}^N T_{ki}  O_{kmi}\big) \big(\sum_{l = 1}^N T_{lj}O_{lmj}\big)}  \\
	 & = \frac{1}{N}\sum_{m = 1}^N \meanb{C_{mi}C_{mj}} = \gamma_{ij},  \\
\end{split}
\]
where $\gamma_{ij} = \frac{1}{N}\sum_{k = 1}^N\mean{C_{ki}C_{kj}}$, and
\[
\begin{split}
	 \sum_{\omega \in \Omega_{i(j + M)}} g(\omega) & = \frac{1}{N}\sum_{k, l, m = 1}^N \mean{T_{ki}O_{kmi}}\mean{T_{lj}O_{lmj}} \\
	 & = \frac{1}{N}\sum_{m = 1}^N \meanb{\sum_{k = 1}^N T_{ki}  O_{kmi}}\meanb{\sum_{l = 1}^N T_{lj}O_{lmj}} \\
	 & = \frac{1}{N}\sum_{m = 1}^N \meanb{C_{mi}}\meanb{C_{mj}} = \mu^2\quad.  \\
\end{split}
\]
The last equality follows from the fact that $\frac{1}{N}\sum_{k = 1}^N C_{kl} = U$. Since $\mean{C_{kl}}$
does not depend on $k$, this immediately implies that $\mean{C_{kl}} = \mean{U} = \mu$.

Let $1 < i \leq M$, Then
\[
\begin{split}
	 \sum_{\omega \in \Omega_{i(M + 1)}} f(\omega) & = \frac{1}{N}\sum_{k, l = 1}^N \mean{T_{ki}O_{kli}T_l} \\
	                                               & = \frac{1}{N}\sum_{k, l, m = 1}^N \mean{T_{ki}O_{kli}T_{li}O_{lmi}} \\
\end{split}
\]
and similarly
\[
\begin{split}
	 \sum_{\omega \in \Omega_{1(M + i)}} f(\omega) & = \frac{1}{N}\sum_{k, l = 1}^N \mean{T_{li}O_{lki}T_k} \\
	                                               & = \frac{1}{N}\sum_{k, l, m = 1}^N \mean{T_{li}O_{lki}T_{ki}O_{kmi}}\quad. \\
\end{split}
\]
Since $O_{kli}O_{lmi} + O_{lki}O_{kmi} = O_{kmi}O_{lmi}$ for $k \neq l$ and $0$ for $k = l$, summing
two previous sums leads to
\[
\begin{split}
	&\frac{1}{N}\sum_{k, l, m = 1}^N \mean{T_{ki}O_{kli}T_{li}O_{lmi} + T_{li}O_{lki}T_{ki}O_{kmi}}  \\
	&\quad = \frac{1}{N}\sum_{\mathrlap{k, l, m = 1}}^N \mean{T_{ki}T_{li}O_{lmi}O_{kmi}} - \frac{1}{N}\sum_{\mathrlap{k, m = 1}}^N \mean{T_{ki}^2O_{kmi}^2} \\
	&\quad = \gamma_{ii}^2 - \frac{1}{N(N - 1)}\sum_{\mathrlap{k = 1}}^N \mean{T_{ki}T_k} \quad.  \\
\end{split}
\]
The last term goes to $0$ as $N$ approaches infinity.
Hence we have
\[
	 \sum_{\omega \in \Omega_{i(M + 1)}} f(\omega) +  \sum_{\omega \in \Omega_{1(M + i)}} f(\omega) \to \gamma_{ii}^2\quad.
\]
On the other hand,
\[
\begin{split}
	 \sum_{\omega \in \Omega_{i(M + 1)}} g(\omega) & = \frac{1}{N}\sum_{k, l = 1}^N \mean{T_{ki}O_{kli}}\mean{T_l}  \\
	 & = \frac{1}{N}\sum_{k}^N \mean{T_k}\mu = \mu^2\quad. \\
\end{split}
\]
and a similar result holds for $\Omega_{1(M + i)}$. Combining all these equations proves the proposition.
\end{IEEEproof}

\subsection{Proof of Proposition~\ref{prop:partstats}} 

\begin{IEEEproof}
Since the blocks $P_i$ are independent, it follows immediately that $\mu = \mean{U} = \prod_{i = 1}^L \mu_i$.
In order to prove the result for the variance, 
let $T_k = \rank{k; X, \mathcal{Y}}$ and $T_{ki} = \rank{i; P_i, \mathcal{Y}}$. 
Let us define
\[
	\alpha = \mean{T_1^2}, \gamma^{(N)} = \mean{T_1T_2}, \gamma_i^{(N)} = \mean{T_{1i}T_{2i}}\quad.
\]
We see that
$\alpha = \prod_{i = 1}^L \alpha_i$ and
$\gamma^{(N)} = \prod_{i = 1}^L \gamma_i^{(N)}$.

Let us define
\[
	\beta_i^{(N)} = \varb{\sqrt{N}U_i} \quad\text{and}\quad \beta^{(N)} = \varb{\sqrt{N}U}\quad.
\]

A straightforward calculation reveals that
\[
	\beta_i^{(N)} = \alpha_i + (N - 1)\gamma_i^{(N)} - N\mu_i^2
\]
and
\[
	\beta^{(N)} = \alpha + (N - 1)\gamma^{(N)} - N\mu^2\quad.
\]

We can express the variance $\beta^{(N)}$ as
\[
\begin{split}
	\beta^{(N)} & = \alpha + (N - 1) \prod_{i = 1}^L \frac{\beta_i^{(N)} - \alpha_i + N\mu_i^2}{N - 1} - N\mu^2\\
	            & = \alpha + \frac{ \prod_{i = 1}^L(\beta_i^{(N)} - \alpha_i + N\mu_i^2) - N(N - 1)^{L - 1}\mu^2}{(N - 1)^{L - 1}}.
\end{split}
\]
Let us now consider the right-hand side as a function of $N$. Both terms in the numerator contain $\mu^2N^L$, consequently
this term is annihilated and the highest term in the numerator has degree of $L - 1$, its
coefficient is equal to
\[
	c_N = (L - 1)\mu^2 + \mu^2\sum_{i = 1}^L \frac{1}{\mu_i^2} (\beta_i^{N} - \alpha_i)\quad.
\]
Since the highest term in the demoninator is $N^{L - 1}$, the fraction converges to $\lim_{N \to \infty} c_N$.
\end{IEEEproof}

\end{document}